# Genomics-guided drawing of malignant regulatory signatures revealed a pivotal role of human stem cell-associated retroviral sequences (SCARS) and functionally-active hESC enhancers.

Gennadi V. Glinsky[1,2]

[1] Institute of Engineering in Medicine

University of California, San Diego

9500 Gilman Dr. MC 0435

La Jolla, CA 92093-0435, USA

[2]OncoSCAR, Inc.

Portland, OR 97207

Correspondence: gglinskii@ucsd.edu

Web: http://iem.ucsd.edu/people/profiles/guennadi-v-glinskii.html

## Abstract

From patients' and physicians' perspectives, the clinical definition of a tumor's malignant phenotype could be restricted to the early diagnosis of sub-types of malignancies with the increased risk of existing therapy failure and high likelihood of death from cancer. It is the viewpoint from which the understanding of malignant regulatory signatures is considered in this contribution. Analyses from this perspective of experimental and clinical observations revealed the pivotal role of human stem cell-associated retroviral sequences (SCARS) in the origin and pathophysiology of clinically-lethal malignancies. SCARS represent evolutionary- and biologically-related family of genomic regulatory sequences, the principal physiological function of which is to create and maintain the stemness phenotype during human preimplantation embryogenesis. SCARS expression must be silenced during cellular differentiation and SCARS activity remains silent in most terminally-differentiated human cells performing specialized functions in the human body. De-repression and sustained activation of SCARS result in differentiation-defective phenotypes, tissue- and organ-specific clinical manifestations of which are diagnosed as pathological conditions defined by a consensus of pathomorphological, molecular, and genetic examinations as the malignant growth. Contemporary evidence are presented that high-fidelity molecular signals of continuing activities of SCARS in association with genomic regulatory networks of thousands' functionally-active enhancers triggering engagements of down-stream genetic loci may serve as both reliable diagnostic tools and druggable molecular targets readily amenable for diagnosis and efficient therapeutic management of clinically-lethal malignancies.

## Preface

The idea that malignant growth originates from stem cells is more than a quarter century old [1]. It was revived at the beginning of 21st century as the cancer stem cell theory [2; 3], which became one of dominant concepts of the contemporary cancer research. One of the key principles of the cancer stem cell theory is that a single cancer stem cell is sufficient to regrow a malignant tumor fully recapitulating morphological, molecular, genomic, and biological features of the parental tumor. Consequently, the theory predicts that cancer cannot be eradicated unless cancer stem cell-targeting therapies [4] will eliminate all cancer stem cells. This postulate is believe to be true because if even a single cancer stem cell would escape the therapeutic assault, it will continue to fuel the malignant growth. However, some fundamental clinical realities seem not necessarily fully compatible with the uniformly simplistic view of the human cancer origin and pathogenesis. First, tumors arising in the same organ are not equivalent in clinical responses to therapies, which could be correlated to their genetic and molecular features. Second, the clinical prognosis related to the organ of cancer origin is markedly different for cancers diagnosed in different organs even at the early stages. Third, in many instances, the clinical cure of malignant tumors has been achieved by the first-line cancer therapies, which are not specifically designed to target cancer stem cells.

On a parallel track, technological advances enabled genome-wide gene expression profiling analyses of human malignancies making a reality the search for gene expression signatures of clinically-lethal malignancies, thus, looking for statistically-significant gene expression correlates of increased likelihood of existing therapy failure and death from cancer. Historically, the theory defining a genomic link between degrees to which a malignancy recapitulates gene expression profiles of stem cells and clinical phenotypes of increased likelihood of therapy failure and death from cancer is originated from the discovery of the death-from-cancer gene expression signature [5]. This genomic connectivity between the phenotypes of resemblance to stemness and high likelihood of death from cancer was initially documented for cancer patients diagnosed with 12 distinct types of human malignancies [5]. Observations reported in the original contributions [5; 6] and follow-up studies (reviewed in 7) directly implicated sustained activation of the Polycomb Group (PcG) Proteins chromatin silencing pathway [8], specifically, the *BMI1* genes, as the principal genomic contributor defining these associations [5; 7]. Collectively, these observations formed the foundation for a concept stating that malignant clinical behaviors of human cancers are governed by stemness genomic laws [5-7; 9-12]. The universal nature of the genomic connectivity between the degree of resemblance to stemness and the extent of malignant behavior of a tumor was validated in numerous experimental cancer models, including transgenic mouse models facilitating implementation of the mouse/human translational genomics approach [5; 13; 14]; clinically-relevant orthotopic xenograft models of human cancers and xenograft-derived cancer cell lines, including blood-borne metastasis precursor cells [6; 15-19]. Mechanistic roles of genes essential for functional integrity of PcG chromatin silencing pathway were demonstrated using targeted genetic interference approaches [13; 20] and gene-specific small molecule therapeutics [21]. Overall, multiple studies have shown that BMI1 inhibition confer therapeutic effects on glioblastoma multioforme, colorectal and breast cancers, as well as chemoresistant ovarian, prostate, pancreatic, and skin cancers [21-24].

However, the major limitation of these and many other early studies was the lack of sufficient understanding of the genomic and molecular underpinning of the stemness phenotype as it emerges during human preimplantation embryogenesis. Remarkable advances in single cell expression profiling analyses of human preimplantation embryos closed this knowledge gap and provided the opportunity to address this limitation. Collectively, these advances facilitated the discovery of stem cell-associated retroviral sequences, which act as the master genomic regulatory elements driving the creation of stemness phenotype in human embryos and may be responsible for its emergence in sub-sets of human malignancies diagnosed in multiple organs.

**Functional, structural, genetic, and molecular definitions of stem cell-associated retroviral sequences (SCARS).**

The term stem cell-associated retroviral sequences (SCARS) refers to the defined set of genomic regulatory sequences sustained expression of which is essential for acquisition and maintenance of stemness phenotype [7; 25; 26]. The canonical definition of "stemness" in reference to human Embryonic Stem Cells (hESC) implies a combination of the phenotypic features of immortality/self-renewal/asymmetrical division/pluripotency. Single cell expression profiling-guided deconvolution of a developmental timeline of human preimplantation embryos enabled the discovery of human embryonic Multi-Lineage Markers Expressing cells (MLME cells), emergence of which during human embryogenesis precedes lineage segregation events and subsequent creation of hESC [27]. Specific members of SCARS termed human pluripotency-associated transcripts (HPATs) have been implicated in the creation of the MLME cells [27]. It has been hypothesized that definition of the "stemness" phenotype for the human MLME cells should be expanded to include the totipotency feature and the human MLME cells could be defined biologically as the pan-lineage precursor cells [27].

For cell differentiation to occur, the expression of SCARS must be silenced: hESC fails to properly differentiate in response to differentiation-inducing cues if SCARS expression is maintained and resulting cells display differentiation-defective phenotypes [28; 29]. It has been suggested that de-repression and sustained re-activation of SCARS expression in association with continuous activation of down-stream genomic regulatory targets (collectively defined as activation of SCARS-associated genomic regulatory networks) is the hallmark of therapy-resistant clinically-lethal malignancies with clinical phenotypes of increased risk of therapy failure and high likelihood of death from cancer [7; 25; 26]. Evolutionary, SCARS are belong to the exceedingly large class of genomic sequences originated from transposable elements (TE) and comprising nearly two-third of human genome. Specifically, in hESC and human preimplantation embryos SCARS represent a functionally-related and structurally-defined sub-set of TE-derived regulatory sequences originated from LTR7/HERV-H, LTR5_Hs/HERV-K, and recently implicated SVA-D retrotransposons [30-32], the set of which was further narrowed by restrictions to human-specific (unique-to-humans) genomic regulatory sequences [7; 25; 26; 33-39].

A range of genetic and molecular definitions of SCARS directly linked to a stemness state extends to different classes of regulatory DNA sequences (transcription factor-binding sites; functional enhancer elements; alternative promoters), donors of splicing sites, and non-coding RNA molecules. Precise mapping of individual transcriptionally-active genomic loci which generated RNA molecules from repetitive sequences (repeats), including highly diverse families of transposable elements (TE) and human endogenous retroviruses (HERV) - encoded sequences, became possible only recently. Advances in RNAseq technology and bioinformatics approaches to data retrieval, processing, and analyses, including implementation of *de novo* transcriptome assembly protocols, facilitated identification of hundreds thousands of TE-encoded RNA molecules precisely mapped to corresponding transcriptionally active genomic loci in human dorsolateral prefrontal cortex [40] and across the spectrum of all major human cancer types [41]. Using the pan-cancer *de novo* transcript assembly approach, the remarkable complexity and ubiquitous nature of transcripts encoded by endogenous retroviral elements (EREs) were uncovered in human malignancies of distinct origins and diverse spectrum of anatomical locations [41]. It has been reported that thousands of transcripts overlapping with regulatory long terminal repeats (LTRs) derived from endogenous retroviruses were expressed in a cancer-specific manner in at least one or several related cancer types [41]. Several of these cancer-specific LTR-harboring transcripts represent relatively large RNA molecules exceeding 50K nucleotides, perhaps, reflecting the read-through transcriptional activity in cancer cells due to the extensive chromatin reprogramming. Notably, cancer-specific RNA molecules derived from individual SCARS loci representing LTR7/HERV-H and LTR5_Hs/HERV-K families accounted for 31% of all reported cancer-specific LTR element-overlapping transcripts that are expressed in more than one cancer type. These cancer-specific LTR-harboring RNA molecules appear to affect the expression of disease-relevant genes and to produce previously unknown cancer-specific antigenic peptides [41]. Therefore, it is now feasible to unequivocally map SCARS-harboring RNA molecules to specific transcriptionally-active genetic loci encoding these transcripts.

**Global DNA methylation reprogramming and SCARS activity contribute to creation of telomerase-positive MLME cells during human preimplantation embryogenesis.**

One of the principal molecular functions of activated SCARS is illustrated by their biological activities attributed to non-coding RNA (ncRNA) molecules transcribed from regulatory DNA segments harboring SCARS. Importantly, manifestations SCARS biological activities have been demonstrated for ncRNAs derived from individual genomic loci [27; 42; 43] and in human embryos SCARS activity has been associated with the creation of telomerase-positive cells co-expressing genetic markers of all embryonic lineages [27]. These telomerase-positive Multi-Lineage Markers Expressing (MLME) cells have been identified employing single cell expression profiling analyses of viable human blastocysts and hundreds of individual cells recovered from preimplantation human embryos [27; 42; 43]. Creation of cells in part resembling gene expression features of MLME cells was recapitulated in genetic engineering experiments, in which individual SCARS-encoded RNAs termed Human Pluripotency-Associated Transcripts (HPATs) were over-expressed in human cells [27; 42; 43]. These observations support the hypothesis that SCARS activation in human embryos may have contributed to the creation of MLME cells.

The summary of the multi-step validation protocol of human embryonic Multi-Lineage Markers Expressing (MLME) cells is shown in the Table 1. The MLME phenotype was assigned to individual telomerase-positive cells that co-expressed at least 6 genetic markers of the Epiblast (EPI) lineage; 7 genetic markers of the Trophectoderm (TE) lineage; and 4 genetic markers of the Primitive endoderm (PE) lineage; and cells must express all three main master pluripotency transcription factors (*OCT4, NANOG, SOX2*). First, the expression levels of 58 genetic markers of human embryonic lineages were considered individually in a particular single cell by comparing the expression values of the markers in a given cell and the median expression value of the marker in the population of single cells of human embryos as previously reported [27; 44]. The marker was considered expressed when the expression value in a cell exceeds the median expression value. The discovery set of 58 genetic markers of human embryonic lineages was utilized in these experiments and based on the above criteria a total of 135 MLME cells were selected from 839 telomerase-positive human embryonic cells. The discovery set of 58 genetic markers of human embryonic lineages was reported elsewhere [27; 45; 46]. Next, independent sets of lineage-specific markers comprising of top 100 individual genetic markers for each embryonic lineage that were utilized for validation of the MLME phenotype in each individually-selected cell. The validation sets of lineage-specific genetic markers of human embryonic lineages were reported elsewhere [44]. To assess the statistical significance of the enrichment of the lineage-specific genetic markers in the MLME cells, p values were estimated using the hypergeometric distribution test. Results of these analyses revealed statistically significant enrichment of genes representing genetic markers of three main embryonic lineages among genes up-regulated in human embryonic MLME cells (Table 1). In agreement with the hypothesis that activities of SCARS contribute to creation of MLME cells, SCARS appear to affect expression of two-third of genes (8,374 of 12,735 genes; 66%) expression of which distinguishes MLME cells from other cells in preimplantation human embryos. Notably, SCARS activity affects expression of a dominant majority (84.1%) of genes up-regulated in human embryonic MLME cells, while expression of only a minor fraction of genes down-regulated in MLME cells (13.4%) appears affected by SCARS.

Zygote-to-embryo transition is accompanied by dramatic DNA methylation reprogramming which is governed by the placeholder nucleosome positioning [47]. Newly established genome-wide dynamics of the chromatin accessibility landscape and concurrent changes of promoter methylation states affect expression of thousands genes and results in embryonic genome activation [31; 48]. Importantly, DNase I hypersensitive site (DHS) sequencing revealed that human transposons SVA and HERV-K harbor DHSs and are highly expressed in early human embryos, but not in differentiated tissues [31]. Analyses of genes comprising GES of human embryonic MLME cells revealed that DNA methylation reprogramming may have contributed to the creation and maintenance of the MLME phenotype in human preimplantation embryos (Figure 1). Collectively, observed in MLME cells gene expression changes of methyltransferases would cause marked reprogramming of genome-wide DNA methylation profiles by erasing the pre-existing cytosine methyl marks and establishing *de novo* methylation patterns (Figure 1A). Concurrently diminished expression of genes encoding primate-specific zinc finger proteins, in particular, *ZNF534* and *ZNF91* genes, would relieve the repressive chromatin from SCARS loci and facilitate activation of SCARS expression (Figure 1B). Consistently, during transition from the oocyte to the morula stage of human preimplantation embryogenesis, promoters of genes comprising the MLME GES shift from nearly exclusively homogenously closed (silenced) states to predominantly homogenously open (active) states (Figure 2). The predominantly homogenously open promoter states of genes comprising the MLME signature are maintained in human embryonic cells of the ICM, TE, and hESC (Figure 2).

**SCARS represent both intrinsic and integral components of human-specific genomic regulatory networks.**

SCARS are predominantly primate-specific regulatory sequences that is they are common for Modern Humans and non-human primates [7]. However, sizable fractions of different SCARS families are represented by human-specific (unique-to-human) regulatory sequences. For example, 302 of 1222 (24.7%) full-length LTR7/HERV-H elements have been identified as candidate human-specific regulatory sequences, HSRS [7]. Species-specificity of SCARS is defined by the unique genomic coordinates of the insertions of corresponding parent transposons, which appear as segments of DNA present on human chromosomes and absent on chromosomes of non-human primates. Interestingly, 37.6% of highly active in hESC LTR7/HERV-H elements have been classified as HSRS [7]. This is contrast to only 19.8% LTR7/HERV-H that are inactive in hESC being identified as candidate HSRS ($p < 0.0001$). Therefore, globally SCARS should be viewed within the genomic regulatory context of other classes of HSRS [38].

Candidate HSRS comprise a coherent compendium of nearly one hundred thousand genomic regulatory elements, including 59,732 HSRS which are markedly distinct in their structure, function, and evolutionary origin [38] as well as 35,074 human-specific neuro-regulatory single nucleotide changes (hsSNCs) located in differentially-accessible (DA) chromatin regions during human brain development [39; 49]. Unified activities of HSRS may have contributed to development and manifestation of thousands human-specific phenotypic traits [39]. SCARS encoded by human endogenous retroviruses LTR7/HERV-H and LTR5_Hs/HERV-K as one of the significant sources of the evolutionary origin of HSRS [7; 25-27; 33-40], including human-specific transcription factor binding sites (TFBS) for NANOG, OCT4, and CTCF [33; 37]. It was interest to determine whether genes previously linked to multiple classes of HSRS, which were identified without considerations of genes expression of which is regulated by SCARS, overlap with SCARS-regulated genes. To this end, 13,824 genes associated with different classes of HSRS were identified using the GREAT algorithm [38; 39], subjected to the GSEA, and compared with the sets of SCARS-regulated genes (Figure 3) identified by shRNA interference [50] and CRISR/Cas-guided epigenetic silencing experiments comparing regulatory networks of naïve and primed hESC [30; 32]. These analyses revealed that SCARS appear to affect expression of a majority (8,384 genes; 61%) of genes associated with different classes of HSRS (Table 2; Supplemental Table S1), in agreement with the hypothesis that a large fraction of SCAS-regulated genes represents an intrinsic component of human-specific genomic regulatory networks. Consistently, SCARS affect expression of a majority of genes (5,389 of 8,405 genes; 64%) associated with neuro-regulatory hsSNCs [39]. Overall, the common gene set of regulatory targets independently defined for HSRS, SCARS, and neuro-regulatory hsSNCs comprises of 7,990 coding genes or 95% of all genes associated with neuro-regulatory hsSNCs located in DA chromatin regions during human brain development [39].

Genes associated with HSRS and neuro-regulatory hsSNCs manifest a staggering breadth of significant associations with morphological structures, physiological processes, and pathological conditions of Modern Humans [39], indicating that a preponderance of human-specific traits evolved under a combinatorial regulatory control of HSRS and neuro-regulatory loci harboring hsSNCs. SCARS-regulated genes comprise a large fraction of these human-specific genomic regulatory networks and represent an integral component of genomic regulatory wiring governing human-specific features of early embryonic development.

One of the important questions is whether the patterns of significant associations with physiological and pathological phenotypes observed for genes linked with HSRS, hsSNCs, and SCARS are specific and not related to the size effects of relatively large gene sets subjected to the GSEA [39]. To address this questions, 42,847 human genes not linked by the GREAT algorithm with HSRS were randomly split into 21 control gene sets of various sizes ranging from 2,847 to 6,847 genes and subjected to the GSEA [39]. Importantly, no significant phenotypic associations were observed for 21 control gene sets, indicating phenotypic associations attributed to genes linked with HSRS, hsSNCs, and SCARS are not likely due to non-specific gene set size effects captured by the GSEA. These observations are highly consistent with the conclusion that a broad spectrum of significant phenotypic associations documented for genes linked with HSRS, neuro-regulatory hsSNCs, and SCARS reflects their bona fide impacts on physiological and pathological phenotypes of Modern Humans. It should be underscored that the efficient execution of these analytical experiments was greatly facilitated by the web-based utilities provided by the Enrichr Bioinformatics System Biology platform [51; 52].

*Gene set enrichment analyses (GSEA) of 8,384 genes associated with HSRS, expression of which is regulated by LTR7Y/B and LTR5_Hs/SVA_D enhancers and HERVH lncRNAs.*

GSEA on multiple genomics databases revealed remarkable breadth and depth of significant associations with physiological and pathological phenotypes of Modern Humans of 8,834 SCARS-regulated genes linked with multiple families of HSRS (Supplemental Text S1). Consistent with the established role of SCARS in human embryogenesis, SCARS-regulated genes are significantly enriched in human embryo and neuronal epithelium according to GSEA of the ARCHS4 Human Tissues database. Consistently, POU5F1 and PRDM14 master stem cell regulators were identified by GSEA of the ESCAPE stem cell-focused database as top up-stream regulators, while pathways in Cancer (KEGG 2019 Human database) and Axon Guidance (KEGG 2019 Mouse database) were scored as top significantly-enriched pathways.

GSEA of the Allan Brain Atlas database focused on up-regulated genes identified 590 human brain regions among significantly enriched records, while GSEA of the Allen Brain Atlas of down-regulated genes identified 847 significant records (adjusted p-value < 0.05). Notably, seven of the top ten significantly enriched records among up-regulated genes identified the Dentate Gyrus, while remaining 3 of the top 10 records identified the Fields CA3 of stratum pyramidale and stratum lucidum of the hippocampus (Supplemental Text S1; Allan Brain Atlas database; up-regulated genes).

GSEA of the Virus MINT database comprising of human genes that encode proteins known to physically interact with viruses and viral proteins identified the Epstein-Barr virus as the top-scoring record, indicating that upon entry in human cells the Epstein-Barr virus-encoded proteins target proteins encoded by SCARS-regulated genes. Overall, expression of nearly 60% of all human genes encoding virus-interacting proteins (2,574 of 4,433 VIP-encoding genes; 58%) is regulated by SCARS.

*GSEA of 2,846 genes associated with created de novo HSRS, expression of which is regulated by LTR7Y/B and LTR5_Hs/SVA_D enhancers and HERVH lncRNAs.*

In human genome, there are 4,528 genes comprising putative regulatory targets of ~12,000 created *de novo* HSRS [38; 39]. Notably, SCARS regulate expression of 2,846 genes (63%) of all genes identified as candidate regulatory targets of created *de novo* HSRS. GSEA of genomics databases revealed numerous significant enrichment records linked with 2,846 SCARS-regulated genes, thus highlighting their potential impacts on human physiology and pathology (Supplemental Text S2).

Unexpectedly, GSEA of the ENCODE and ChEA Consensus transcription factors (TFs) from ChIP-X database identified androgen receptor (AR) as a top-scoring candidate upstream regulator. In agreement with the above observations, GSEA of the ARCHS4 Human Tissues database identified Neuronal epithelium, Human embryo, and Prefrontal cortex as top significantly-enriched records (Supplemental Text S2). Pathways in Cancer (KEGG 2019 Human database) and Axon Guidance (KEGG 2019 Mouse database) were identified as top significantly enriched pathways. Additionally, pathways of Integrins in angiogenesis (NCI-Nature 2016 database) and Integrin signaling (Panther 2016 database) were identified as top-scoring significantly-enriched pathways (Supplemental Text S2).

GSEA of the Jensen Tissues database identified 134 significantly enriched records indicating that SCARS-regulated genes associated with created *de novo* HSRS have been previously identified among genes comprising expression signatures of many human tissues. Other notable findings were revealed by the GSEA of the Human Phenotype Ontology database (81 significant records); the MGI Mammalian Phenotype 2017 database (309 significant records); the Allen Brain Atlas databases of up-regulated genes (284 significantly-enriched brain regions) and down-regulated genes (408 significantly-enriched brain regions).

SCARS-regulated genes appear significantly enriched among genes implicated in a broad spectrum of human common and rare diseases. GSEA of the Rare Diseases AutoRIF ARCHS4 Predictions database captured 353 significantly-enriched records of human rare disorders (Supplemental Text S2). GSEA of the Disease Perturbations from Gene Expression Omnibus (GEO) database of up-regulated genes identified 246 significant records, while interrogation of the Disease Perturbations from GEO database of down-regulated genes revealed 203 significantly-enriched records (Supplemental Text S2).

Lastly, according to GSEA of the Jensen Diseases database, a significant majority of SCARS-regulated genes associated with created *de novo* HSGRS (2008 of 2846 genes; 71%) have been implicated in development and clinical manifestations of multiple types of human cancers (Supplemental Text S2).

**Inference of potential impacts of SCARS on development and clinical behavior of human malignancies.**

SCARS activation hypothesis postulates the central role of a sustained activity of SCARS in acquisition and maintenance of stemness features in human cancer cells, clinical manifestations of which are reflected in high likelihood of therapy failure and death from cancer [7; 25; 26]. This intrinsic propensity to evade the malignancy eradication therapies is proposed to exist even if SCARS-activation driven cancer is diagnosed as the early stage disease based on established pathomorphological and molecular criteria.

Observations capturing the principal molecular, genetic, and biological features attributed to regulatory impacts of SCARS were made in experimental models of naïve and primed hESC, human induced pluripotent stem cells (iPSC), and human preimplantation embryogenesis. These experiments identified genes expression of which is significantly altered in human cells subjected to targeted genetic manipulations to achieve SCARS over-expression [27; 42; 43} and/or silencing using shRNA interference [42; 43; 50], CRISPR/Cas gene knockout technology [42] as well as CRISPR/Cas-guided epigenetic silencing of SCARS [32], thus facilitating identification of multiple gene expression signatures (GES) reflecting fine details of experimentally-defined SCARS-associated genomic regulatory networks.

*Impacts of genes comprising distinct GES regulated by LTR7Y/B and LTR5_Hs/SVA_D enhancers and HERVH lncRNAs.*

Potential biological relevance of several experimentally-defined GES comprising distinct panels of SCARS-regulated genes have been evaluated using Gene Set Enrichment Analyses (GSEA) across multiple genomic databases as previously described [38; 39]. These analytical experiments were executed using the web-based tools of the Enrichr Bioinformatics System Biology platform [51; 52]. To date, the following GES of SCARS-regulated networks in hESC are available for follow-up interrogations of their biological impacts and potential translational significance:

1. GES comprising a set of 1,141 genes that are regulated by both HERVH lncRNA and LTR5_Hs/SVA_D enhancers;
2. GES comprising a set of 3,063 genes regulated by both LTR7Y/B enhancers and HERVH lncRNA;
3. GES comprising a set of 1,477 genes regulated by both LTR7Y/B enhancers and HERVH lncRNA and manifesting concordant expression profiles;

4. GES comprising a set of 1,586 genes regulated by both LTR7Y/B enhancers and HERVH lncRNA and manifesting discordant expression profiles;

The up to date summary of the key findings for each of these four SCARS GES is reported in the Supplemental Text S3. Notably, GSEA of 1,141 genes that are regulated by both LTR5_Hs/SVA_D enhancers and HERV-H lncRNA facilitated identification and characterization of sub-sets of SCARS-regulated genes implicated in Parkinson's, autism, multiple types of cancer, and human embryonic development (Supplemental Text S3).

GSEA of the Jensen Diseases database revealed that a significant majority of genes regulated by both HERV-H lncRNA and LTR7Y/B enhancers (1905 of 3063 genes; 62%) have been implicated in development and clinical manifestations of multiple types of human cancer. Similarly, a significant majority of genes regulated by both HERV-H lncRNA and LTR7Y/B enhancers and manifesting concordant expression profiles (972 of 1477 genes; 66%) have been implicated in development and clinical manifestations of multiple types of cancer (Supplemental Text S3).

**HSRS and SCARS regulate expression of a majority of cancer survival predictor genes and cancer driver genes.**

One of the approaches to evaluation of potential impacts of SCARS on development and clinical manifestations of human malignancies could be the assessment of regulatory effects of SCARS on cancer survival and cancer driver genes. To this end, analyses of 10,713 protein-coding genes expression changes of which are significantly associated with the increased likelihood of survival of cancer patients diagnosed with 17 major cancer types [53] and 460 cancer driver genes identified in 28 human cancer types [54] revealed that SCARS regulate a majority of either cancer survival genes or cancer driver genes (Tables 3; 4; Figure 4; Supplemental Text S4).

It has been observed (Table 3) that a prominent majority of human cancer survival predictor genes is regulated by HSRS (7,738 genes; 72%). As shown in Table 4, SCARS regulate expression of 7,609 of 10,713 (71%) human cancer survival predictor genes (Table 4).

SCARS regulate expression of two-third cancer driver genes (305 of 460 genes; 66%) and as many as 73-75% of high-confidence cancer driver genes (Figure 4), which were defined by either the level of peer-reviewed literature support (Figure 1A) or the statistical significance levels (Figure 4B). Notably, SCARS regulate expression of a majority of cancer driver genes regardless of their maximum mutations' frequency (Figure 4D). SCARS-regulated cancer driver genes were identified in all analyzed to date 28 types of human cancer (Table 5). From the therapeutic strategy stand point, it is important to map actionable cancer therapy-guiding nodes defined by the SCARS stemness matrix mapped to connect Cancer Driver Genes/Cancer Type/Regulatory SCARS (Table 5). Further details describing regulatory effects of HSRS and SCARS on cancer survival and cancer driver genes are reported in the Supplemental Text S4.

**Implications for mechanistic studies of normal development and pathophysiology of Modern Humans.**

One of the most intriguing molecular functions of SCARS is highlighted by their role as functionally active enhancers as well as the ability of SCARS to influence enhancers' activity. Enhancer elements could be divided into functionally silent and functionally active categories. Exceedingly large set of functionally silent enhancers could be defined by characteristic chromatin marks indicating that specific DNA sequences harboring these chromatin marks may function as enhancer elements. Accurate molecular and genetic definitions of functionally active enhancers require the application of specific assays in a particular cell type as it has been reported for hESC [55]. It has been observed that SCARS are significantly enriched among regulatory DNA sequences identified in either primed or naïve hESC as functionally active enhancer elements [37; 55]. Furthermore, human embryonic MLME cells, creation of which was associated with SCARS activity [27], appear to capture GES of both Naïve and Primed hESC (Supplemental Text S5) with more significant resemblance of hESC in the Naïve state. Therefore, assessments of biological roles of functionally active enhancers in hESC may shed a light on our understanding of potential biological impacts of SCARS-associated genomic regulatory networks.

Arguably, two key biologically-distinct functions of active enhancers in hESC are the maintenance of self-renewal and pluripotency states by restricting the differentiation potential and changing on demand the expression of genes linked to major embryonic lineages. Primed hESCs, in particular, are thought to represent a state poised to differentiation in which functionally active enhancers linked to differentiation of various lineages can be quickly switched on or off in response to developmental cues (likely in response to changes in chromatin and histone modification patterns). The biological role of functionally active hESC enhancers could be inferred by evaluating the enrichment within regulatory networks governed by naïve and primed hESC enhancers of genes comprising expression signatures of different human and non-human embryonic lineages (Table 6). In these analyses gene expression signatures of major embryonic lineages of distinct species, including humans, monkeys, and mice were evaluated [27; 45; 46; 50; 56-58]. To this end, all genes comprising expression signatures of distinct embryonic lineages were assessed and genes which are located in close genomic proximity (at a distance of 10 kb or less) to naïve and primed hESC functionally active enhancers were identified. It has been observed that in all instances a high proportion of marker genes distinguishing embryonic lineages are located in close genomic proximity to hESC functional enhancers (Table 6). Notably, proportions of genes associated with naïve and primed hESC enhancers appear similar, consistent with the hypothesis that both naive and primed hESC represent functionally distinct states with the complimentary relevance to mechanistic exploration of developmental pathways.

To assess the statistical significance of these findings, observed numbers of genes associated with hESC functional enhancers were compared to the expected values based on associations by chance alone. Expected values were estimated based on the number of genes in the human genome (63,677); number of genes associated with functional enhancers of the Naïve hESC (18,766); number of genes associated with functional enhancers of the Primed hESC (17,131); number of genes associated with functional enhancers of both Naive and Primed hESC (25,421); and numbers of genes in the corresponding expression signatures of embryonic lineages. These analyses revealed that in all instances differences between the observed and expected numbers of observations appear highly statistically significant (Table 6). These findings indicate that genomic networks governed by both naïve and primed functional enhancers in hESC may represent valuable models for follow-up mechanistic studies of regulatory mechanisms governing critical stages of the human pre-implantation embryogenesis. This line of investigations have been extended to evaluate the potential biological role of hESC functionally active enhancers by performing the proximity placement analyses of genes associated with regulatory networks of naïve and primed hESC functional enhancers and compare these with genes involved in human embryonic, neurodevelopmental, and cancer survival predictors' transcriptional networks (Supplemental Tables S9 and S10), which were previously identified in multiple independent studies {27; 30; 40; 45; 46; 50; 53; 56-60]. A comprehensive genome-wide proximity placement analyses identifies all genes associated with functional enhancers, which were defined based on the location of their genomic coordinates within +/- 10 Kb windows of the corresponding enhancer’s genomic coordinates [55]. All genes in common have been identified for a set of genes associated with enhancers and a set of genes comprising the expression signatures of corresponding embryonic, neurodevelopmental, and cancer survival predictors' networks. Finally, the assessment of statistical significance of observed versus expected numbers of genes in common has been performed for corresponding gene sets. Highly significant associations (Supplemental Tables S9 and S10) of genes defining human embryonic, neurodevelopmental, and cancer survival predictors' transcriptional networks with naïve (Supplemental Tables S9) and primed (Supplemental Tables S10) hESC functionally active enhancers have been observed. Genes associated with functionally active enhancers in Naïve and Primed hESC are significantly enriched for genes comprising human-specific expression signatures of excitatory neurons (Figure 5A), radial glia (Figure 5B), induced pluripotent cells (Figure 5C), and human genes encoding a majority of virus-interacting proteins (Figure 5D). It should be noted that these regulatory genomic features of functionally active hESC enhancers are markedly similar to the regulatory impacts of HSRS and SCARS on genes implicated in pathogenesis of neurodevelopmental, neuropsychiatric, and neurodegenerative disorders [35; 38; 39]. The summary of observations supporting this conclusion is reported in the Supplemental Text S6.

Collectively, these findings strongly argue that a comprehensive catalog of functionally active enhancers in hESC together with GES of SCARS-regulated genes may serve as an important previously unavailable resource for evidence-based mechanistic dissections of fine genomic regulatory architectures governing expression of genes implicated in transcriptional networks relevant to human development and diseases.

**Evolutionary aspects of the emergence of overlapping genetic networks associated with cancer and other common human disorders.**

Present analyses support the idea of shared genomic regulatory networks impacting human cancer survival, neuropsychiatric, neurodevelopmental, and neurodegenerative disorders. Many genes that expressed in the human brain tend to be long because they have more introns, which is also true for genes expressed in specific cells in human preimplantation embryos because there is a large overlapping genetic networks operating in MLME cells of human embryos and fetal/adult neocortex of human brains. Overall, we have more introns in our genes than, for example mouse, and about 10% less protein coding genes. Thus, in genomes of Modern Humans high transcripts' diversity (which impacts both regulatory diversity of RNA molecules and diversity of peptides and proteins) was achieved by inserting more intronic sequences and increasingly relying on splicing. Retrotransposition is one of the major mechanistic contributors to these continuing processes with major impacts on stem cells survival and expansion to sustain the regeneration and dying cells' replenishment in various tissues and organs (Figure 6). DNA sequences of long genes that are expressed and continuously transcribed in these long living cells for many years of the individuals' lifetime have a significantly higher probability to acquire and accumulate functionally deleterious, regulatory, and disease causing mutations. Depending on when and where it happened, it would manifest as different diseases: for example, in cells of coherent peripheral tissues it would be diagnosed as malignant tumors, while in cells of central nervous system it would be diagnosed as neurodevelopmental, neuropsychiatric, or neurodegenerative disorders. It has been suggested [33] that, in addition to deamination of methyl-cytosine, one of the main mechanisms promoting the increased likelihood of mutations is the RNA-mediated formation of energetically-stable DNA:RNA triple-stranded complexes designated R-loops, specifically, R-loops formation of which is driven by SCARS-encoded RNA molecules.

**Hypothesis of a singular source code for essential faithful execution of early embryogenesis programs and driving the emergence of disease states in human cells.**

Precisely controlled waves of activities of distinct families of transposable elements, including SCARS, provides a genomic source code for proper execution of high-complexity developmental programs during human preimplantation embryogenesis. In human embryonic stem cells (hESC), sustained activities of SCARS is required for maintenance of the stemness state. Conversely, failure to silence SCARS during neuronal differentiation of hESC is associated with development of differentiation-defective phenotypes, indicating that SCARS activity is not compatible with physiological functions of differentiated human cells. Consequently, aberrant activation of SCARS in long-living human cells might represent a genomic source code driving the emergence of various disease states, including cancer, neurodegeneration, neurodevelopmental and neuropsychiatric disorders (Figure 6). In this contribution, experimental evidence and theoretical considerations have been summarized supporting the model of a singular genomic source code, activation and execution of which contributes to development of multiple types of human disorders. This singular genomic source code captures the mechanistic essence of malignant regulatory signatures.

1. Removing of DNA and chromatin silencing marks at SCARS-encoding loci.
2. Activation of transcription at SCARS loci and production of SCARS RNAs.
3. Activation of a genome-wide network of functional enhancers associated with a pluripotent state.
4. Genome-scale effects of functional enhancers and SCARS-encoded RNAs on DNA conformation dynamics and chromatin states resulting in expression changes of thousands genes, including known specific disease state-causing genes.
5. Effects of SCARS-encoded RNAs on functions and maintenance of microRNAs affecting the stability, abundance, and translation of mRNAs.
6. Amplification of transcriptional activation of transposable elements, including SVA, Alu, and LINE retrotransposons.
7. Formation of energetically stable triple-stranded RNA/DNA complexes transitioning to highly prone to mutations R-loop conformations.
8. Binding of SCARS RNAs to conformation-disordered proteins, including RNA-binding proteins and virus interacting proteins (VIPs).
9. Genome-scale rewiring of and interference with liquid-liquid phase-separated condensates affecting chromatin regulatory states.
10. System-scale effects on the efficiency of compartmentalization of biochemical reactions controlled by liquid-liquid phase-separated condensates.

## Conclusions

In accord with the expected in vivo regulatory role of SCARS and hESC functional enhancers during human embryonic development, a significant enrichment of genes comprising expression signatures of major embryonic lineages of distinct species, including humans, monkeys, and mice has been observed within regulatory networks of Naïve and Primed hESC functional enhancers. Results of these analyses further support the hypothesis that key regulatory features of human neurodevelopmental networks are engaged during the early-stages of human embryogenesis [33-35; Supplemental Tables S11-S12]. Analyses of regulatory networks of Naïve and Primed hESC functional enhancers revealed a highly consistent pattern of significant enrichment of genes that were previously identified as principal components of major neurodevelopmental networks (Supplemental Tables S9-S12), including GES of human neuronal and non-neuronal brain cells [59], human neurons' sub-types and neuronal diversity signatures [60], and human fetal brain/adult neocortex GES [35]. Consistent with the idea that activation of stemness genomic networks in cancer cells contributes to development of clinically-lethal death-from-cancer phenotypes, interrogation of regulatory networks of SCARS as well as Naïve and Primed hESC functional enhancers revealed a significant enrichment of cancer survival predictors' genes that were defined for 17 distinct types of human malignancies [53]. Similar regulatory connectivity has been observed for SCARS and cancer driver's genes identified for 28 human cancer types [54]. Importantly, in all instances these analyses demonstrated that regulatory networks of SCARS and functional enhancers operating in hESC in both Naïve and Primed states appear to capture distinct arrays of genomic regulatory networks engaged in human embryogenesis, neurodevelopmental processes, and human malignancies. Consequently, collective considerations of all observations summarized in this contribution strongly argue that highly tractable experimental model systems tailored for precise structure-activity-phenotype interrogations of SCARS and functional enhancers in both Naïve and Primed hESC would represent a valuable, perhaps, indispensable, resource for dissections of principal genetic elements governing primate-specific and unique to human features of development, physiology, and pathology of Modern Humans.

**Perspectives**

- From the clinical perspective, perhaps, reflecting the best interest of cancer patients, the most important translational impact of malignant regulatory signatures would be the reliable early diagnosis of sub-types of malignancies with the increased risk of existing therapy failure and high likelihood of death from cancer. It is this yet unfulfilled promise of malignant regulatory signatures defining stemness of human malignancies is the main focus of this contribution.
- The predominant focus of the contemporary research effort on elucidation of molecular interconnectivity of the stemness phenotype and development of human cancers is on advancement of the cancer stem cell concept. The impact of recent remarkable advancements of single cell genomics of preimplantation human embryos, the bone fide source of the stemness phenotype creation during human development, had relatively modest influence on cancer research and, in particular, on progress in our understanding of mechanistic underpinning of malignant regulatory signatures.
- The in-depth analyses of the critically important impact of stem cell associate retroviral sequences (SCARS) as the essential elements of malignant regulatory signatures of clinically lethal human cancers will be one of the main topic of the future research. These studies should include precise identification and detailed structure-function analyses of individual transcriptionally-active regulatory genomic loci harboring SCARS and down-stream target genes making vital contributions to pathogenesis of human malignancies and multiple other common disorders.

## Methods

### Data source and analytical protocols

A total of 94,806 candidate HSRS, including 35,074 neuro-regulatory human-specific SNCs, detailed descriptions of which and corresponding references of primary original contributions are reported elsewhere (Glinsky et al., 2015-2020; Kanton et al., 2019). Solely publicly available datasets and resources were used in this contribution. The significance of the differences in the expected and observed numbers of events was calculated using two-tailed Fisher's exact test. Additional placement enrichment tests were performed for individual classes of HSRS taking into account the size in bp of corresponding genomic regions. Detailed description of methodological and analytical approaches are provided in the Supplemental Methods and previously reported contributions (Glinsky et al., 2015-2020).

*Statistical Analyses of the Publicly Available Datasets*

All statistical analyses of the publicly available genomic datasets, including error rate estimates, background and technical noise measurements and filtering, feature peak calling, feature selection, assignments of genomic coordinates to the corresponding builds of the reference human genome, and data visualization, were performed exactly as reported in the original publications and associated references linked to the corresponding data visualization tracks (http://genome.ucsc.edu/). Any modifications or new elements of statistical analyses are described in the corresponding sections of the Results. Statistical significance of the Pearson correlation coefficients was determined using GraphPad Prism version 6.00 software. Both nominal and Bonferroni adjusted p values were estimated. The significance of the differences in the numbers of events between the groups was calculated using two-sided Fisher's exact and Chi-square test, and the significance of the overlap between the events was determined using the hypergeometric distribution test (Tavazoie et al., 1999).

**Table 1.** Enrichment of genes comprising top 100 lineage-specific genetic markers of each of three major embryonic lineages of human preimplantation embryos among genes that are significantly up-regulated in the MLME cells.

| Classification category | Number of genes | Number of up-regulated genes in the MLME cells | Percent | P value* | Observed/expected ratio** |
|---|---|---|---|---|---|
| **Human genome** | 26178 | 9430 | 36.0 | | |
| **Genetic markers of the human Epiblast (EPI)** | 100 | 91 | 91.0 | 1.186E-30 | 2.53 |
| **Genetic markers of the human Primitive Endoderm (PE)** | 88 | 41 | 46.6 | 0.0107581 | 1.29 |
| **Genetic markers of the human Trophectoderm (TE)** | 100 | 81 | 81.0 | 2.799E-20 | 2.25 |

Legend: *, p values were estimate using the hypergeometric distribution test; **, expected values were estimated based on the number of all analyzed genes (26178) and the number of genes significantly up-regulated in the human embryonic MLME (9430); MLME, multi-lineage markers expressing cells; A total of 819 telomerase-positive (*TERT*pos) individual human embryonic cells were analyzed and each single cell was identified as the putative immortal MLME cell if it expressed genetic markers of each of the 3 major lineages (epiblast, EPI; throphectoderm, TE; and primitive endoderm, PE) and all three (*NANOG; POU5F1; SOX2*) pluripotent state master regulators. The MLME phenotype was assigned to individual telomerase-positive cells that co-expressed at least 6 genetic markers of the EPI lineage; 7 genetic markers of the TE lineage; and 4 genetic markers of the PE lineage; and three main master pluripotency transcription factors. The expression levels of 58 genetic markers of human embryonic lineages were considered individually in a particular single cell by comparing the expression values of the markers in a given cell and the median expression value of the marker in the population of single cells of human embryos as previously reported (Petropoulos et al., 2016; Glinsky et al., 2018). The marker was considered expressed when the expression value in a cell exceeded the median expression value. The set of 58 genetic markers of human embryonic lineages analyzed in these experiments during the selection a total of 135 MLME cells from 839 TERTpos human embryonic cells is listed in the Supplemental Table S4 (Glinsky et al., 2018) and was originally reported elsewhere (Blakeley et al., 2015).. Independent sets of lineage-specific markers comprising of top 100 individual genetic markers for each embryonic lineage were utilized for validation of the MLME phenotype and were reported elsewhere (Petropoulos et al., 2016).

**Table 2.** SCARS regulate expression of a majority of 13,824 genes associated with human-specific regulatory sequences (HSRS).

| Classification category | Number of genes | Percent* |
|---|---|---|
| **HERV-H lncRNA-regulated genes** | 4805 | 34.76 |
| **LTR7Y/B enhancers-regulated genes** | 5240 | 37.91 |
| **LTR5_Hs/SVA_D enhancers-regulated genes** | 2022 | 14.63 |
| **All SCARS-regulated HSRS-associated genes** | 8384 | 60.65 |

Legend: *, percent of all HSRS-associated genes.

**Table 3.** A prominent majority of human cancer survival predictor genes is associated with human-specific regulatory sequences (HSRS).

| TYPE OF CANCER | CANCER SURVIVAL GENES | HSRS-ASSOCIATED | PERCENT |
|---|---|---|---|
| **Thyroid** | 347 | 269 | 77.52 |
| **Glioma** | 271 | 206 | 76.01 |
| **Melanoma** | 205 | 153 | 74.63 |
| **Head and neck** | 808 | 597 | 73.89 |
| **Colorectal** | 603 | 440 | 72.97 |
| **Renal** | 6070 | 4418 | 72.78 |
| **Ovarian** | 504 | 366 | 72.62 |
| **Liver** | 2892 | 2086 | 72.13 |
| **Lung** | 662 | 477 | 72.05 |
| **Breast** | 582 | 414 | 71.13 |
| **Urothelial** | 1101 | 783 | 71.12 |
| **Stomach** | 307 | 218 | 71.01 |
| **Prostate** | 161 | 114 | 70.81 |
| **Endometrial** | 1631 | 1153 | 70.69 |
| **Cervical** | 717 | 505 | 70.43 |
| **Pancreatic** | 1549 | 1075 | 69.40 |
| **Testis** | 60 | 42 | 70.00 |
| **All human cancer survival genes** | 10713 | 7738 | 72.23 |

Legend: Numbers of genes in each cell reflect the sum of records in the corresponding classification category when individual genes were recorded as a single count. Uhlen et al. (2017) reported a total of 10,713 protein-coding genes expression changes of which are significantly associated with the increased likelihood of survival of cancer patients diagnosed with 17 major cancer types. Percent values were calculated as fractions of all cancer survival genes in the corresponding classification categories.

**Table 4.** SCARS regulate expression of a prominent majority of human cancer survival predictor genes.

| TYPE OF CANCER | CANCER SURVIVAL GENES | SCARS-REGULATED | PERCENT | HERV-H-REGULATED | PERCENT | LTR7Y/B-REGULATED | PERCENT | LTR5_Hs/SVA_D-REGULATED | PERCENT |
|---|---|---|---|---|---|---|---|---|---|
| **BREAST** | 582 | 405 | 69.59 | 229 | 39.35 | 284 | 48.80 | 80 | 13.75 |
| **PROSTATE** | 161 | 121 | 75.16 | 63 | 39.13 | 90 | 55.90 | 10 | 6.21 |
| **PANCREATIC** | 1549 | 1112 | 71.79 | 629 | 40.61 | 772 | 49.84 | 250 | 16.14 |
| **LIVER** | 2892 | 2217 | 76.66 | 1267 | 43.81 | 1565 | 54.11 | 382 | 13.21 |
| **RENAL** | 6070 | 4406 | 72.59 | 2579 | 42.49 | 2881 | 47.46 | 965 | 15.90 |
| **COLORECTAL** | 603 | 448 | 74.30 | 262 | 43.45 | 320 | 53.07 | 104 | 17.25 |
| **CERVICAL** | 717 | 526 | 73.36 | 312 | 43.51 | 340 | 47.42 | 112 | 15.62 |
| **LUNG** | 662 | 488 | 73.72 | 298 | 45.02 | 312 | 47.13 | 105 | 15.86 |
| **THYROID** | 347 | 259 | 74.64 | 153 | 44.09 | 171 | 49.28 | 56 | 16.14 |
| **OVARIAN** | 504 | 368 | 73.02 | 202 | 40.08 | 233 | 46.23 | 78 | 15.48 |
| **ENDOMETRIAL** | 1631 | 1129 | 69.22 | 652 | 39.98 | 747 | 45.80 | 250 | 15.33 |
| **UROTHELIAL** | 1101 | 772 | 70.12 | 458 | 41.60 | 483 | 43.87 | 164 | 14.90 |
| **HEAD & NECK** | 808 | 558 | 69.06 | 340 | 42.08 | 369 | 45.67 | 128 | 15.84 |
| **GLIOMA** | 271 | 204 | 75.28 | 115 | 42.44 | 128 | 47.23 | 48 | 17.71 |
| **MEANOMA** | 205 | 148 | 72.20 | 85 | 41.46 | 107 | 52.20 | 25 | 12.20 |
| **STOMACH** | 307 | 219 | 71.34 | 144 | 46.91 | 131 | 42.67 | 25 | 8.14 |
| **TESTIS** | 60 | 41 | 68.33 | 23 | 38.33 | 26 | 43.33 | 11 | 18.33 |
| **ALL** | 10713 | 7609 | 71.03 | 4436 | 41.41 | 5013 | 46.79 | 1641 | 15.32 |

Legend: Numbers of genes in each cell reflect the sum of records in the corresponding classification category when individual genes were recorded as a single count. Uhlen et al. (2017) reported a total of 10,713 protein-coding genes expression changes of which are significantly associated with the increased likelihood of survival of cancer patients diagnosed with 17 major cancer types. Percent values were calculated as fractions of all cancer survival genes in the corresponding classification categories.

**Table 5.** SCARS-guided cancer stemness matrix of diagnostic and therapeutic targets comprising of 237 SCARS-down-regulated and 141 SCARS-activated cancer driver genes mapped to 28 cancer types.

| Cancer Type | Number of SCARS-silenced cancer driver genes | Number of SCARS-activated cancer driver genes |
|---|---|---|
| **Adenoid Cystic** | 7 | 4 |
| **Bladder** | 30 | 16 |
| **Blood** | 25 | 22 |
| **Brain** | 28 | 16 |
| **Breast** | 28 | 17 |
| **Cervix** | 12 | 8 |
| **Cholangiocarcinoma** | 8 | 4 |
| **Colorectal** | 16 | 12 |
| **Endometrium** | 30 | 20 |
| **Gastroesophageal** | 37 | 26 |
| **Head & Neck** | 19 | 6 |
| **Kidney Clear** | 8 | 5 |
| **Kidney Non-Clear** | 14 | 6 |
| **Liver** | 18 | 10 |
| **Lung AD** | 15 | 9 |
| **Lung SC** | 7 | 3 |
| **Lymph** | 36 | 25 |
| **Ovarian** | 4 | 2 |
| **Pancreas** | 22 | 17 |
| **Pheochromocytoma** | 5 | 3 |
| **Pleura** | 9 | 0 |
| **Prostate** | 19 | 11 |
| **Sarcoma** | 6 | 3 |
| **Skin** | 21 | 13 |
| **Testicular Germ Cell** | 11 | 8 |
| **Thymus** | 7 | 2 |
| **Thyroid** | 8 | 9 |
| **Uveal Melanoma** | 2 | 1 |
| **Number of actionable cancer therapy-guiding nodes defined by the SCARS stemness matrix mapped to connect Cancer Driver Genes/Cancer Type/Regulatory SCARS** | **1365** | **834** |

Legend: AD, adenocarcinoma; SC, small cell carcinoma;

**Table 6.** Enrichment within regulatory networks of Naïve and Primed hESC active enhancers of gene expression signatures (GES) defining embryonic lineages of distinct species.

| **Human epiblast (EPI) vs naïve hESC (hESCp#0)** | | | | |
|---|---|---|---|---|
| **Classification category** | Number of genes | Percent | P value* | Observed/Expected |
| **Human EPI vs hESCp#0 GES** | 1496 | 100.0 | | |
| **Naïve functional enhancers network** | 762 | 50.9 | 1.544E-69 | 1.73 |
| **Primed functional enhancers network** | 726 | 48.5 | 5.669E-73 | 1.80 |
| **Naïve & Primed functional enhancers networks** | 976 | 65.2 | 4.326E-89 | 1.63 |
| **Human epiblast (EPI) vs Trophectoderm (TE) GES** | | | | |
| **Classification category** | Number of genes | Percent | P value* | Observed/Expected |
| **Human EPI vs TE expression signature** | 836 | 100.0 | | |
| **Naïve functional enhancers network** | 525 | 62.8 | 1.176E-89 | 2.13 |
| **Primed functional enhancers network** | 472 | 56.5 | 2.095E-73 | 2.10 |
| **Naïve & Primed functional enhancers networks** | 647 | 77.4 | 2.11E-109 | 1.94 |
| **Monkey epiblast (EPI) GES** | | | | |
| **Classification category** | Number of genes | Percent | P value* | Observed/Expected |
| **Monkey EPI expression signature** | 719 | 100.0 | | |
| **Naïve functional enhancers network** | 442 | 61.5 | 1.665E-71 | 2.09 |
| **Primed functional enhancers network** | 399 | 55.5 | 1.687E-59 | 2.06 |
| **Naïve & Primed functional enhancers networks** | 529 | 73.6 | 1.003E-75 | 1.84 |
| **Mouse inner cell mass (ICM) vs Trophectoderm (TE)** | | | | |
| **Classification category** | Number of genes | Percent | P value* | Observed/Expected |
| **Mouse ICM vs TE expression signature** | 497 | 100.0 | | |
| **Naïve functional enhancers network** | 246 | 49.5 | 2.533E-21 | 1.68 |
| **Primed functional enhancers network** | 211 | 42.5 | 2.314E-14 | 1.58 |
| **Naïve & Primed functional enhancers networks** | 303 | 61.0 | 1.061E-21 | 1.53 |

Legend: *, p values were estimated using the hypergeometric distribution test; expected values were estimated based on the number of genes in the human genome (63,677); number of genes associated with functional enhancers of the Naïve hESC (18,766); number of genes associated with functional enhancers of the Primed hESC (17,131); and number of genes associated with functional enhancers of both Naive and Primed hESC (25,421); GES, gene expression signature;

## Figure legends

**Figure 1.** Expression changes of genes encoding DNA methyltransferases and primate-specific zinc finger proteins in human embryonic MLME cells.

A. Telomerase-positive MLME cells manifest decreased expression of the *DNMT1* gene, which is responsible for genome-wide maintenance of DNA methylation patterns, and increased expression of genes responsible for genome-wide *de novo* methylation patterns (*DNMT3A, DNMT3B, DNMT3L*).

B. Concurrently, MLME cells exhibit decreased expression of primate-specific zinc finger proteins responsible for sequence-specific silencing of SCARS and other TE-harboring loci during human preimplantation embryogenesis. Collectively, these changes of gene expression cause marked reprogramming of DNA methylation patterns in genomes of MLME cells and are associated with activation of SCARS expression.

MLME cells are designated as immortal multi-lineage precursor cells, iMPC [27].

**Figure 2.** Dynamics of promoter state's changes of genes comprising GES of human embryonic MLME cells during human preimplantation embryogenesis.

Graphs reflect the gradual transition from predominantly homogenously closed (silent) promoter state in the oocyte to predominantly homogenously open (active) promoter state at the morula stage. Homogenously open promoter states of genes comprising the MLME GES [27] are maintained in human embryonic cells of the ICM, TE, and hESC. Divergent promoter state definition refers to a transitional state of partially closed and partially open promoters.

Promoter states of human genes at different stages of preimplantation embryogenesis were reported elsewhere [48].

**Figure 3.** Genome-wide gene expression profiling experiments identify thousands of SCARS-regulated genes in hESC.

Genome-wide RNAseq analyses were performed on genetically engineered hESC to identify genes regulated by SCARS-encoded regulatory signals derived from HERV-H, LTR5_Hs/SVA_D, and LTR7Y/B loci. Genes regulated by HERV-H ncRNA molecules were identified using shRNA-mediated genetic interference [50], while genes regulated by LTR5_Hs/SVA_D and LTR7Y/B enhancers were identified employing CRISPR/Cas-guided epigenetic silencing [32].

**Figure 4.** SCARS regulate expression of a prominent majority of cancer driver genes.

A total of 460 cancer driver genes reported in [54] were evaluated for regulatory dependency from SCARS.

A. SCARS regulate expression of a prominent majority of high-confidence cancer driver genes defined by different levels of peer-review literature support.

B. SCARS regulate expression of a prominent majority of high-confidence cancer driver genes defined by different levels of statistical significance.

C. Distinct families of SCARS regulate expression of cancer driver genes collectively affecting expression of a prominent majority of cancer driver genes.

D. SCARS regulate expression of a prominent majority of cancer driver genes defined by different levels of mutation frequency.

E. Direct correlation between numbers of SCARS-activated and SCARS-silenced cancer driver genes in 28 human cancer types.

**Figure 5.** Networks of genes regulated in Naïve and Primed hESC hESC by functionally-active enhancers are enriched for genes comprising human-specific expression signatures of excitatory neurons (A), radial glia (B), induced pluripotent cells (C), and human genes encoding a majority of virus-interacting proteins (D).

**Figure 6.** SCARS pathways – guided control of differentiated cells' replenishment cycles.

A. A model of the cycle in the balanced state.

B. A model of the cycle at the loss of differentiated cells state.

C. A model of the cycle at the completed replenishment of differentiated cells state.

D. A model of the cycle with failed replenishment of differentiated cells due to failure of the SCARS silencing during attempts toward differentiation.

A

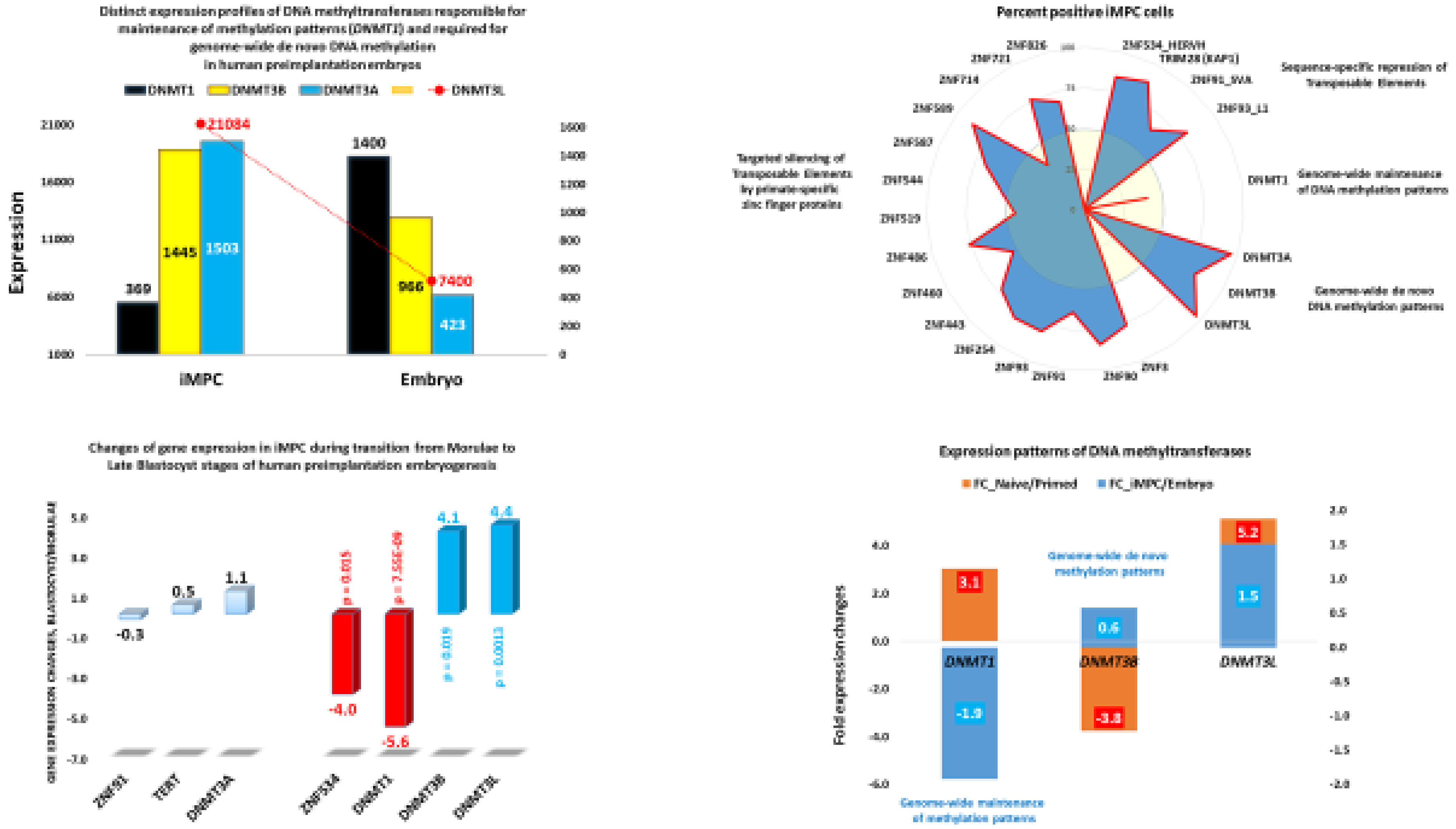
Distinct expression profiles of DNA methyltransferases responsible for maintenance of methylation patterns (DNMT1) and required for genome-wide de novo DNA methylation in human preimplantation embryos
DNMT1
DNMT3B
DNMT3A
DNMT3L
Expression
21084
1400
369
1445
1503
966
7400
423
iMPC
Embryo
Percent positive iMPC cells
Sequence-specific repression of Transposable Elements
Targeted silencing of Transposable Elements by primate-specific zinc finger proteins
Genome-wide maintenance of DNA methylation patterns
Genome-wide de novo DNA methylation patterns
DNMT1
DNMT3A
DNMT3B
DNMT3L
Changes of gene expression in iMPC during transition from Morulae to Late Blastocyst stages of human preimplantation embryogenesis
-0.3
0.5
1.1
-4.0
-5.6
4.1
4.4
ZNF534
DNMT1
DNMT3B
DNMT3L
Expression patterns of DNA methyltransferases
FC_Naive/Primed
FC_iMPC/Embryo
Fold expression changes
3.1
-1.9
0.6
-3.8
5.2
1.5
DNMT1
DNMT3B
DNMT3L
Genome-wide de novo methylation patterns
Genome-wide maintenance of methylation patterns

B

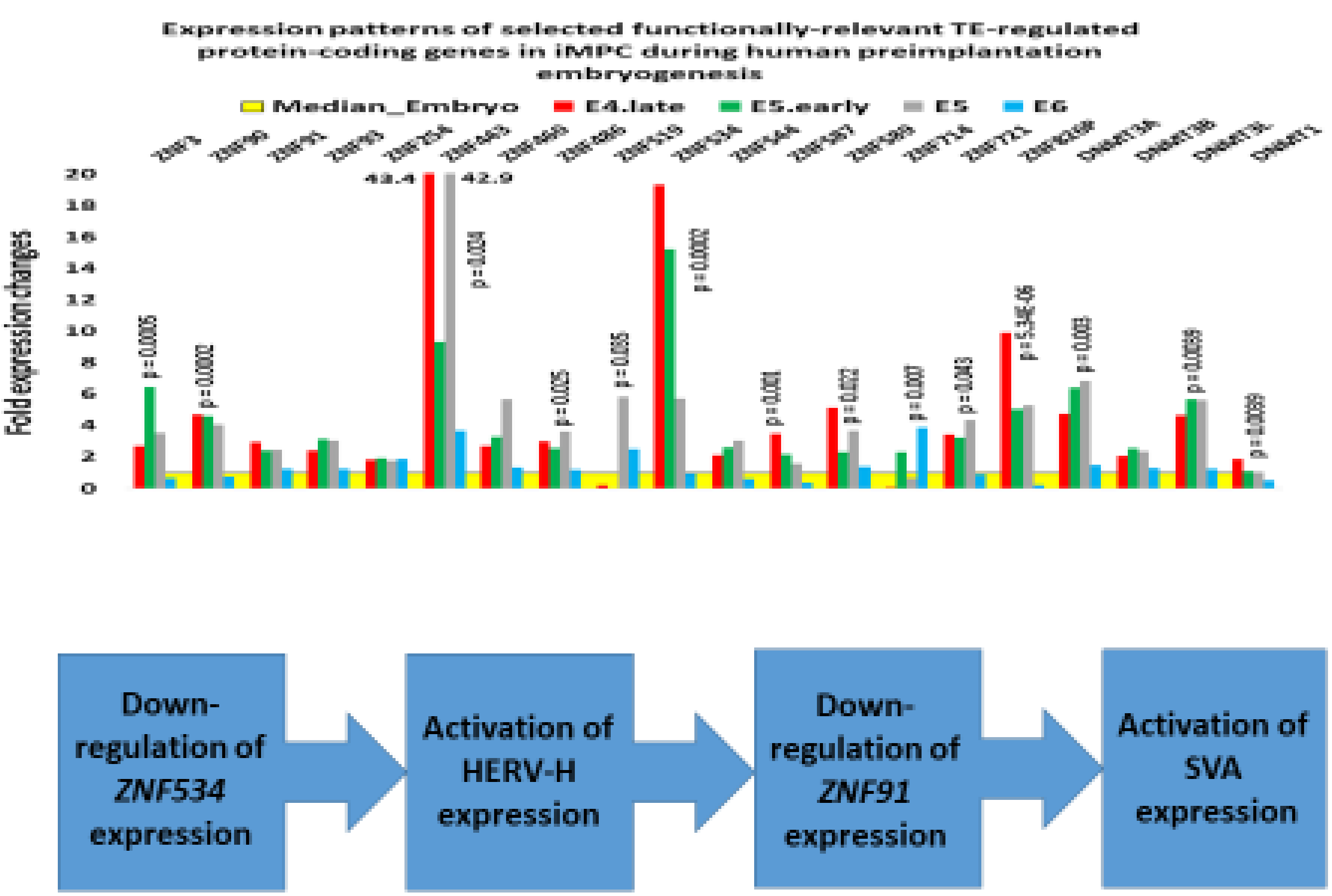

Expression patterns of selected functionally-relevant TE-regulated protein-coding genes in iMPC during human preimplantation embryogenesis
Median_Embryo
E4.late
E5.early
E5
E6
Fold expression changes
42.4
42.9
Down-regulation of ZNF534 expression
Activation of HERV-H expression
Down-regulation of ZNF91 expression
Activation of SVA expression

**FIGURE 1.**

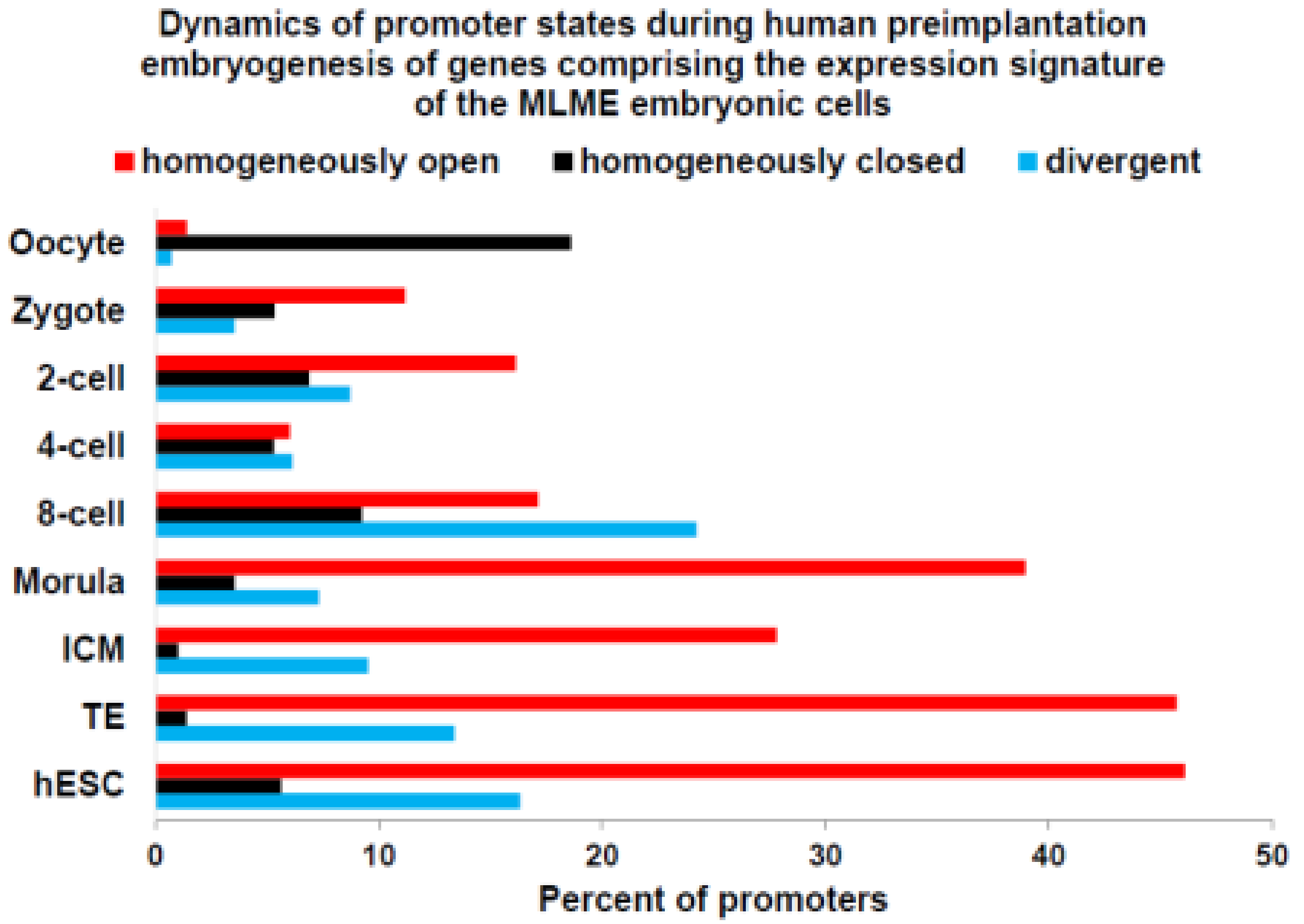
Dynamics of promoter states during human preimplantation embryogenesis of genes comprising the expression signature of the MLME embryonic cells
homogeneously open
homogeneously closed
divergent
Oocyte
Zygote
2-cell
4-cell
8-cell
Morula
ICM
TE
hESC
0
10
20
30
40
50
Percent of promoters

**FIGURE 2.**

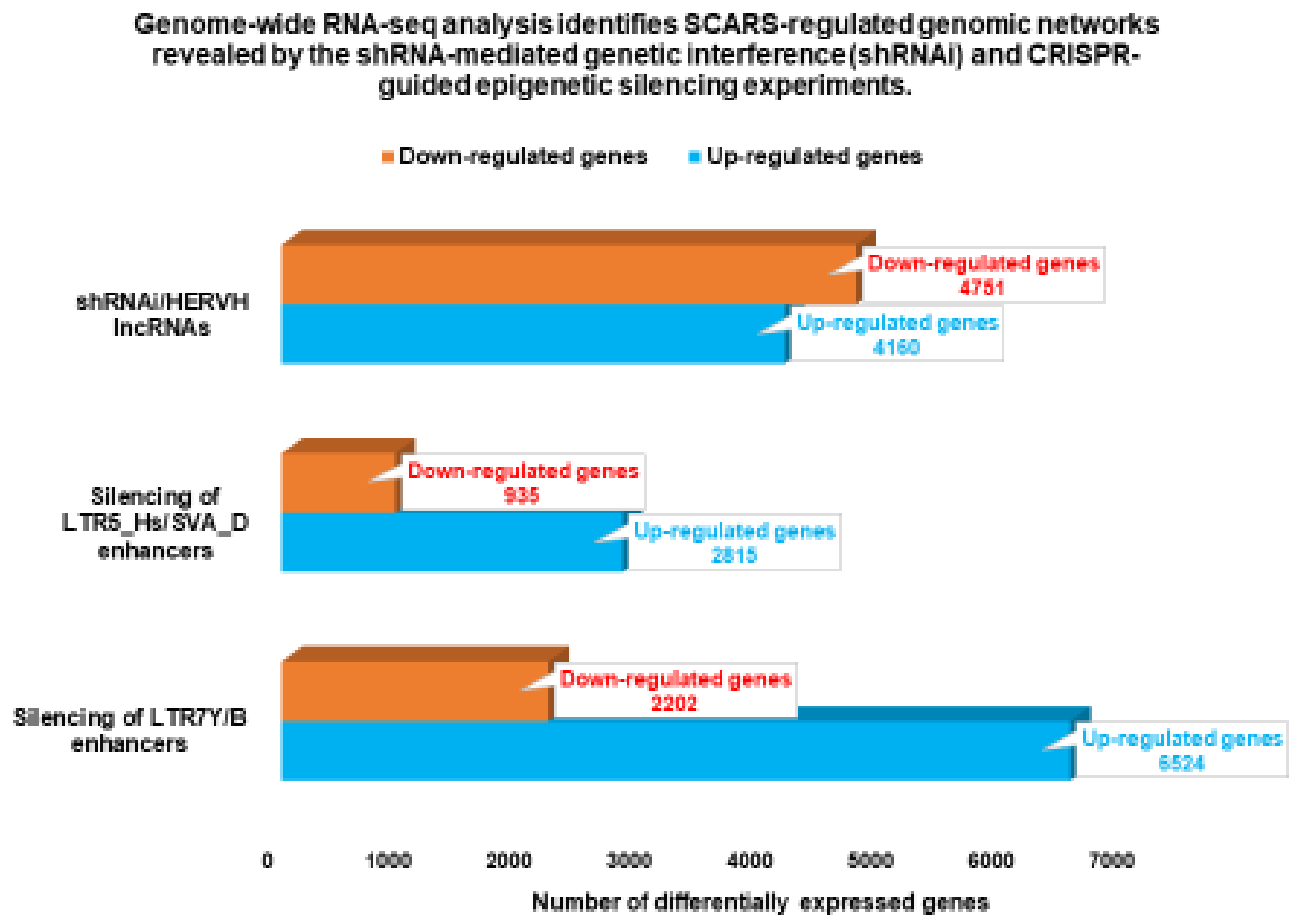
Genome-wide RNA-seq analysis identifies SCARS-regulated genomic networks revealed by the shRNA-mediated genetic interference (shRNAi) and CRISPR-guided epigenetic silencing experiments.
Down-regulated genes
Up-regulated genes
shRNAi/HERVH lncRNAs
Silencing of LTR5_Hs/SVA_D enhancers
Silencing of LTR7Y/B enhancers
Down-regulated genes 4751
Up-regulated genes 4160
Down-regulated genes 935
Up-regulated genes 2815
Down-regulated genes 2202
Up-regulated genes 6524
0
1000
2000
3000
4000
5000
6000
7000
Number of differentially expressed genes

**FIGURE 3.**

A. SCARS regulate expression of a prominent majority of high-confidence cancer driver genes defined by different levels of peer-review literature support

| Category of cancer driver genes | Peer-review literature support level | Number of cancer driver genes | SCARS-regulated cancer driver genes | Percent |
|---|---|---|---|---|
| level A | Genes listed in the Cancer Gene Census | 194 | 145 | 74.74 |
| level B | Genes with literature support in the same tumor types as those identified by NG2020 method | 90 | 62 | 68.89 |
| level C | Genes with literature support in a different tumor type | 98 | 64 | 65.31 |
| level D | Genes with no literature support | 78 | 34 | 43.59 |
| All cancer driver genes | | 460 | 305 | 66.30 |

B. SCARS regulate expression of a prominent majority of high-confidence cancer driver genes defined by different levels of statistical significance

| Minimum FDR cut-off level | Number of cancer driver genes | SCARS-regulated cancer driver genes | Percent |
|---|---|---|---|
| 0 to <0.001 | 178 | 130 | 73.03 |
| >0.001 to <0.05 | 93 | 63 | 67.74 |
| >0.05 to <0.150 | 91 | 58 | 63.74 |
| >0.150 | 98 | 54 | 55.10 |
| All cancer driver genes | 460 | 305 | 66.30 |

C. Distinct families of SCARS regulate expression of cancer driver genes collectively affecting expression of a prominent majority of cancer driver genes

| Classification category | Number of cancer driver genes | Percent of all cancer driver genes |
|---|---|---|
| HERV-H-regulated cancer driver genes | 185 | 40.22 |
| LTR7Y/B-regulated cancer driver genes | 195 | 42.39 |
| LTR5_Hs/SVA_D-regulated cancer driver genes | 91 | 19.78 |
| All SCARS-regulated cancer driver genes | 305 | 66.30 |

D. SCARS regulate expression of a prominent majority of cancer driver genes defined by different levels of mutation frequency

| Max. Mutation Frequency (%) | Number of cancer driver genes | SCARS-regulated cancer driver genes | Percent |
|---|---|---|---|
| > 50% | 6 | 5 | 83.33 |
| >=20% to <50% | 20 | 13 | 65.00 |
| >=10% to <20% | 56 | 40 | 71.43 |
| >=5% to <10% | 93 | 65 | 69.89 |
| >=3% to <5% | 99 | 66 | 66.67 |
| 1% to 2% | 186 | 116 | 62.37 |
| All cancer driver genes | 460 | 305 | 66.30 |

E

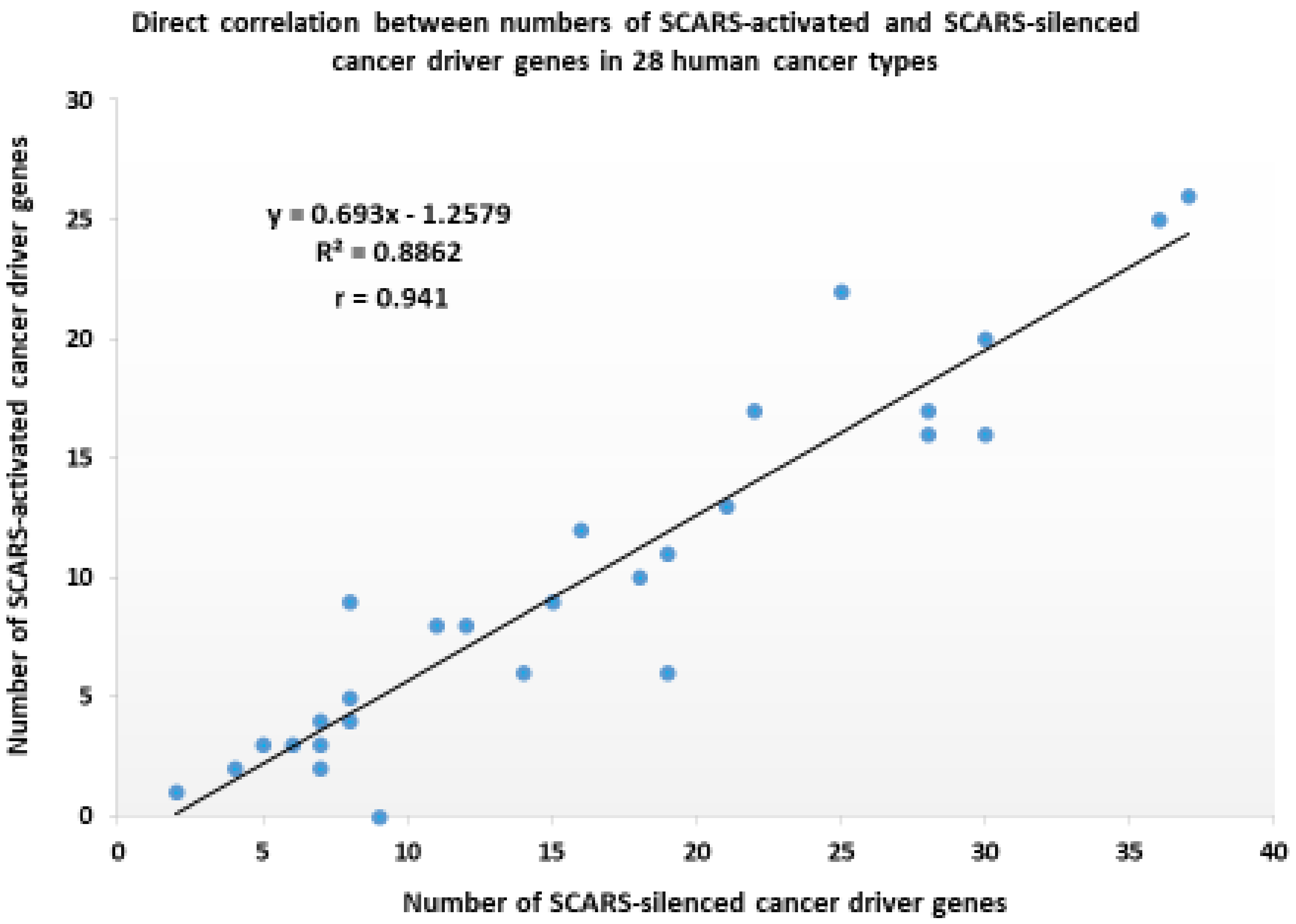

Direct correlation between numbers of SCARS-activated and SCARS-silenced cancer driver genes in 28 human cancer types
y = 0.693x - 1.2579
R² = 0.8862
r = 0.941
Number of SCARS-activated cancer driver genes
0
5
10
15
20
25
30
0
5
10
15
20
25
30
35
40
Number of SCARS-silenced cancer driver genes

**FIGURE 4.**

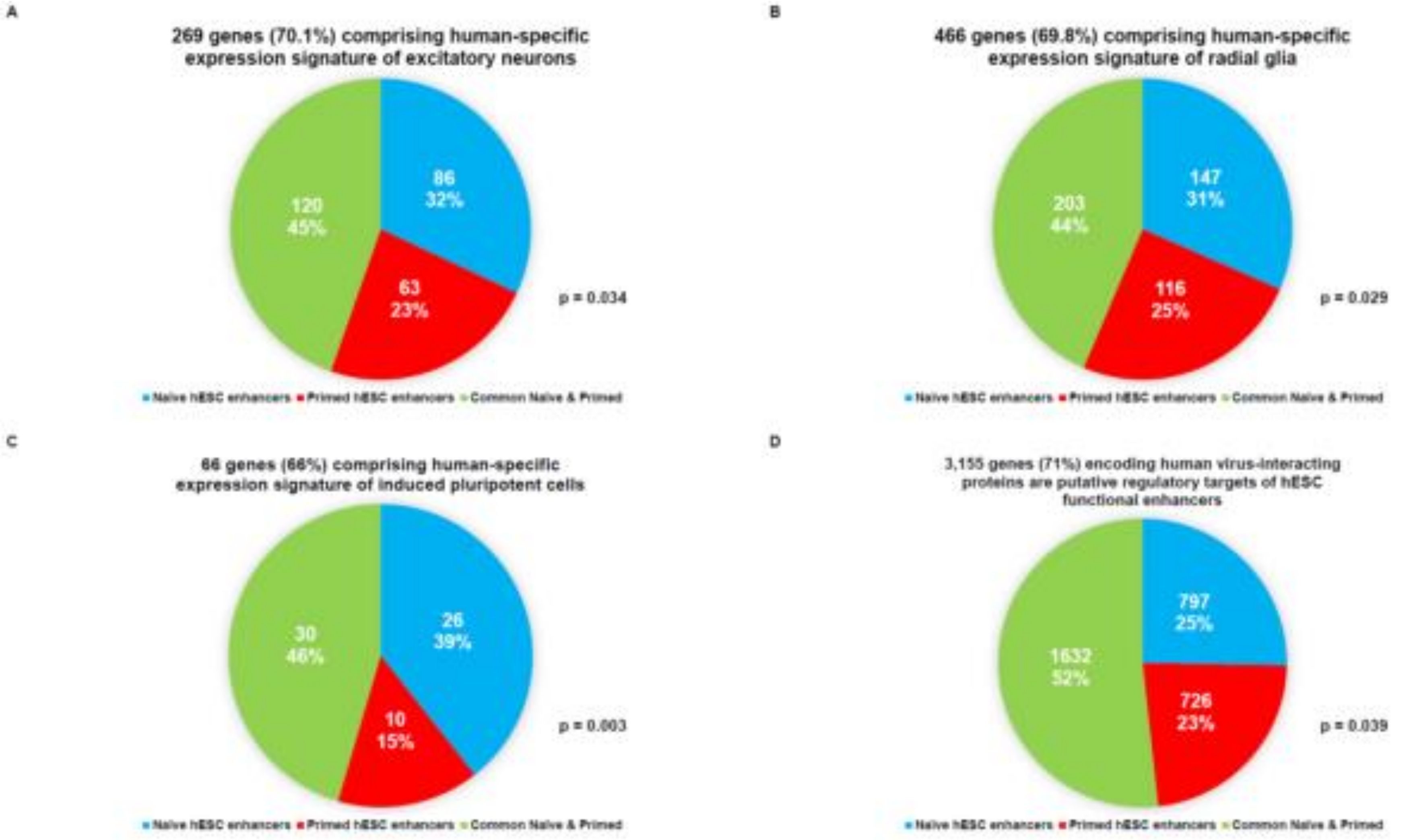
A
269 genes (70.1%) comprising human-specific expression signature of excitatory neurons
86
32%
63
23%
120
45%
p = 0.034
Naive hESC enhancers
Primed hESC enhancers
Common Naive & Primed
B
466 genes (69.8%) comprising human-specific expression signature of radial glia
147
31%
116
25%
203
44%
p = 0.029
Naive hESC enhancers
Primed hESC enhancers
Common Naive & Primed
C
66 genes (66%) comprising human-specific expression signature of induced pluripotent cells
26
39%
10
15%
30
46%
p = 0.003
Naive hESC enhancers
Primed hESC enhancers
Common Naive & Primed
D
3,155 genes (71%) encoding human virus-interacting proteins are putative regulatory targets of hESC functional enhancers
797
25%
726
23%
1632
52%
p = 0.039
Naive hESC enhancers
Primed hESC enhancers
Common Naive & Primed

**FIGURE 5.**

A

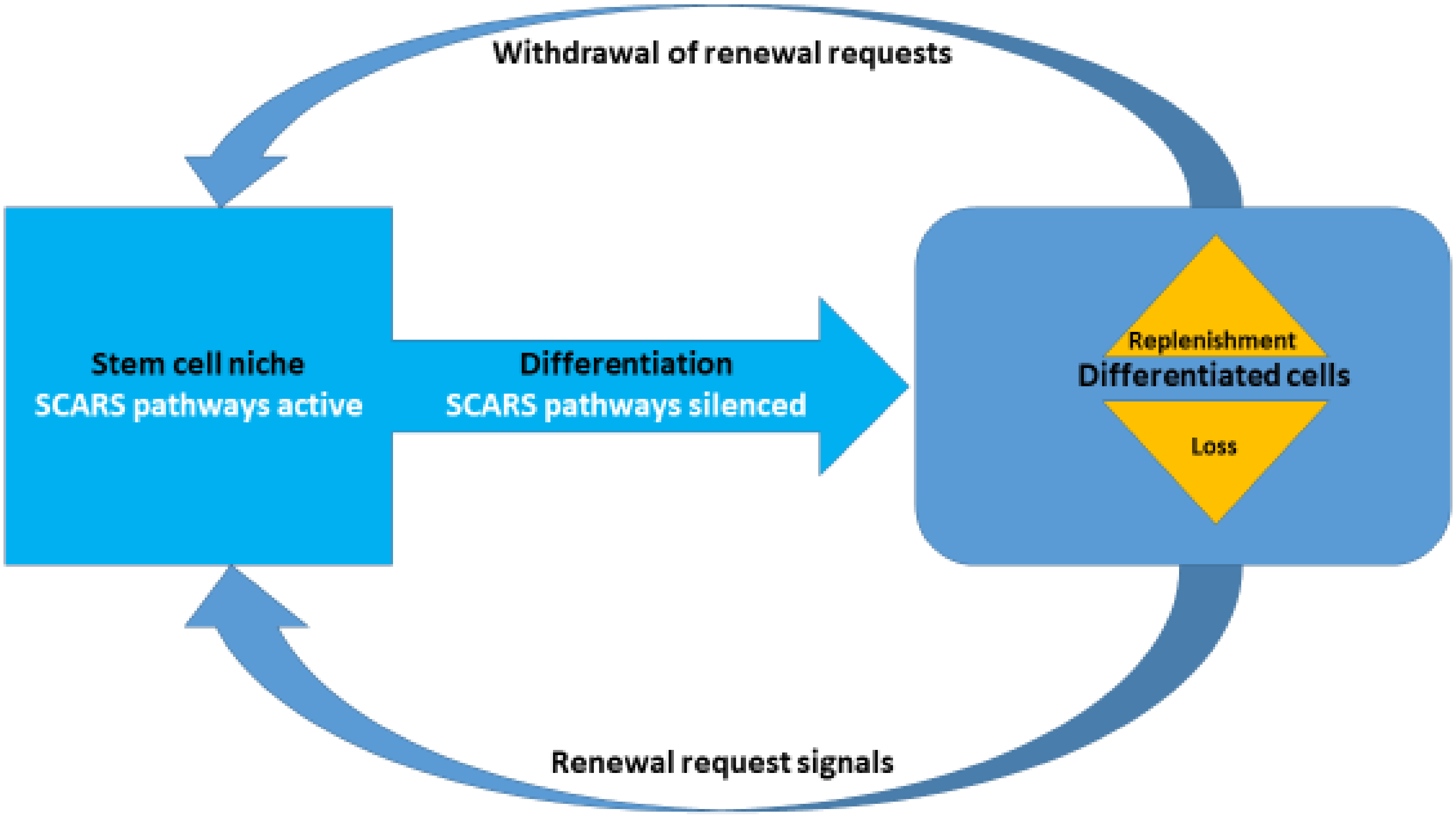

B

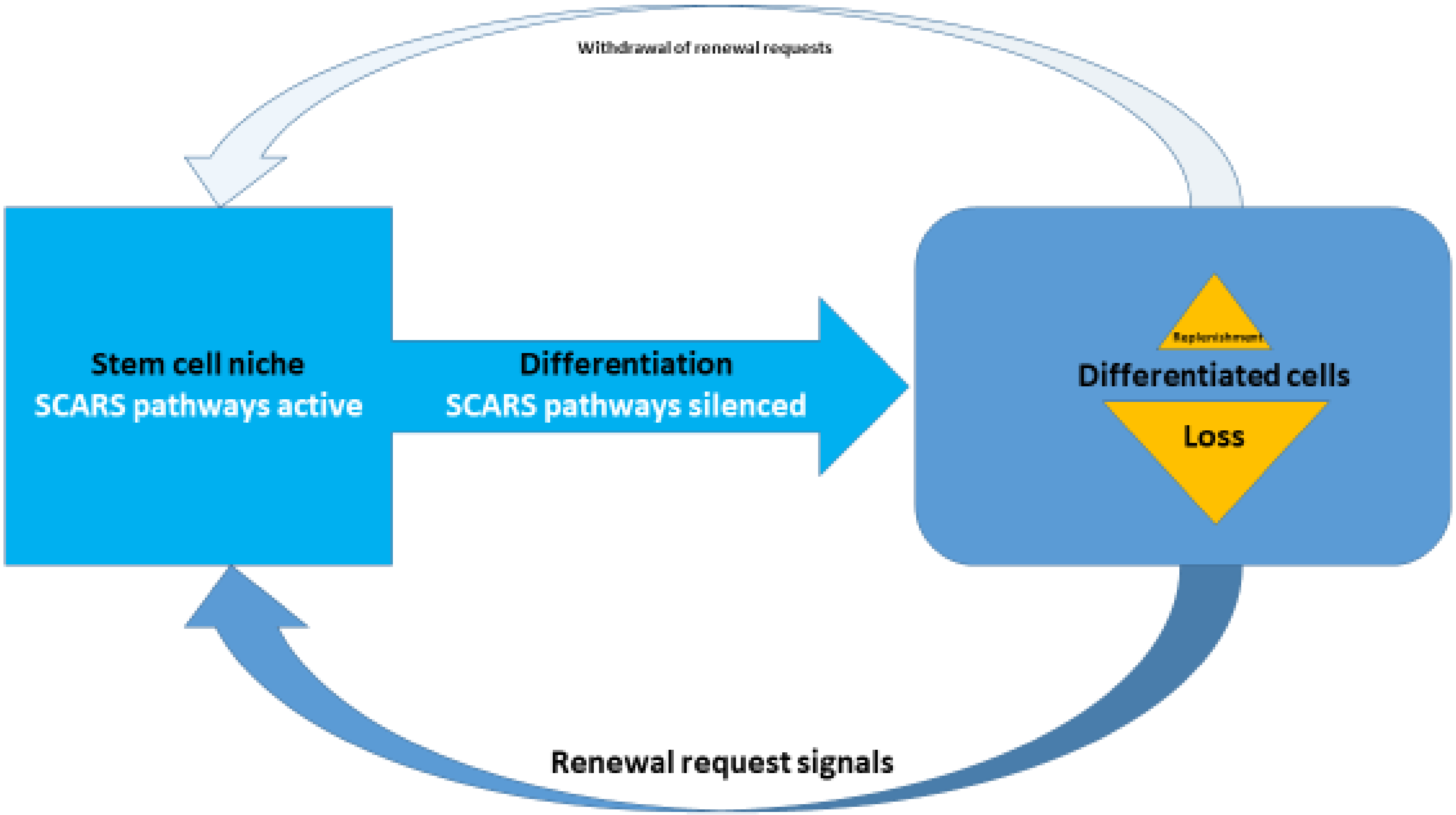

C

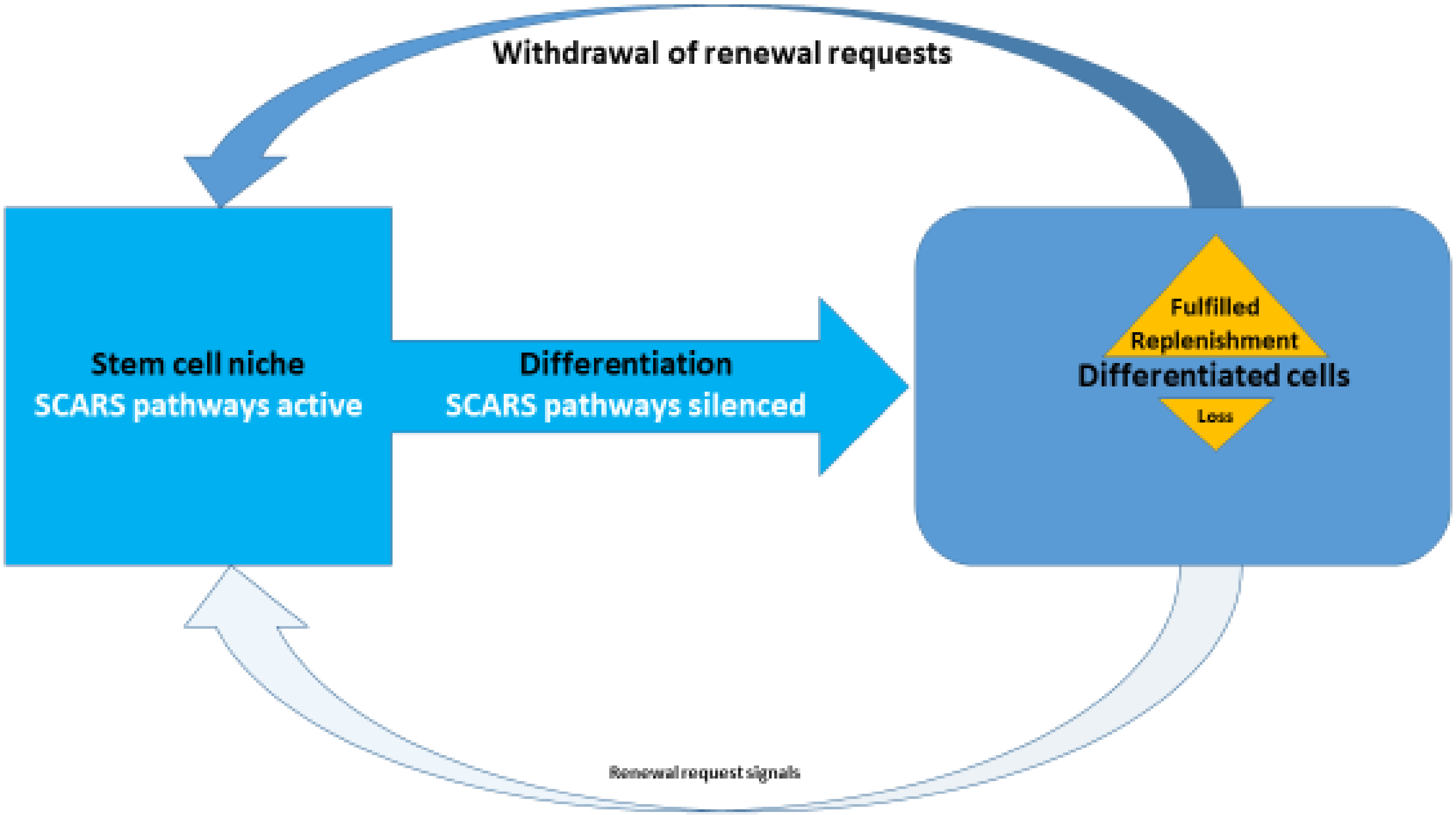

D

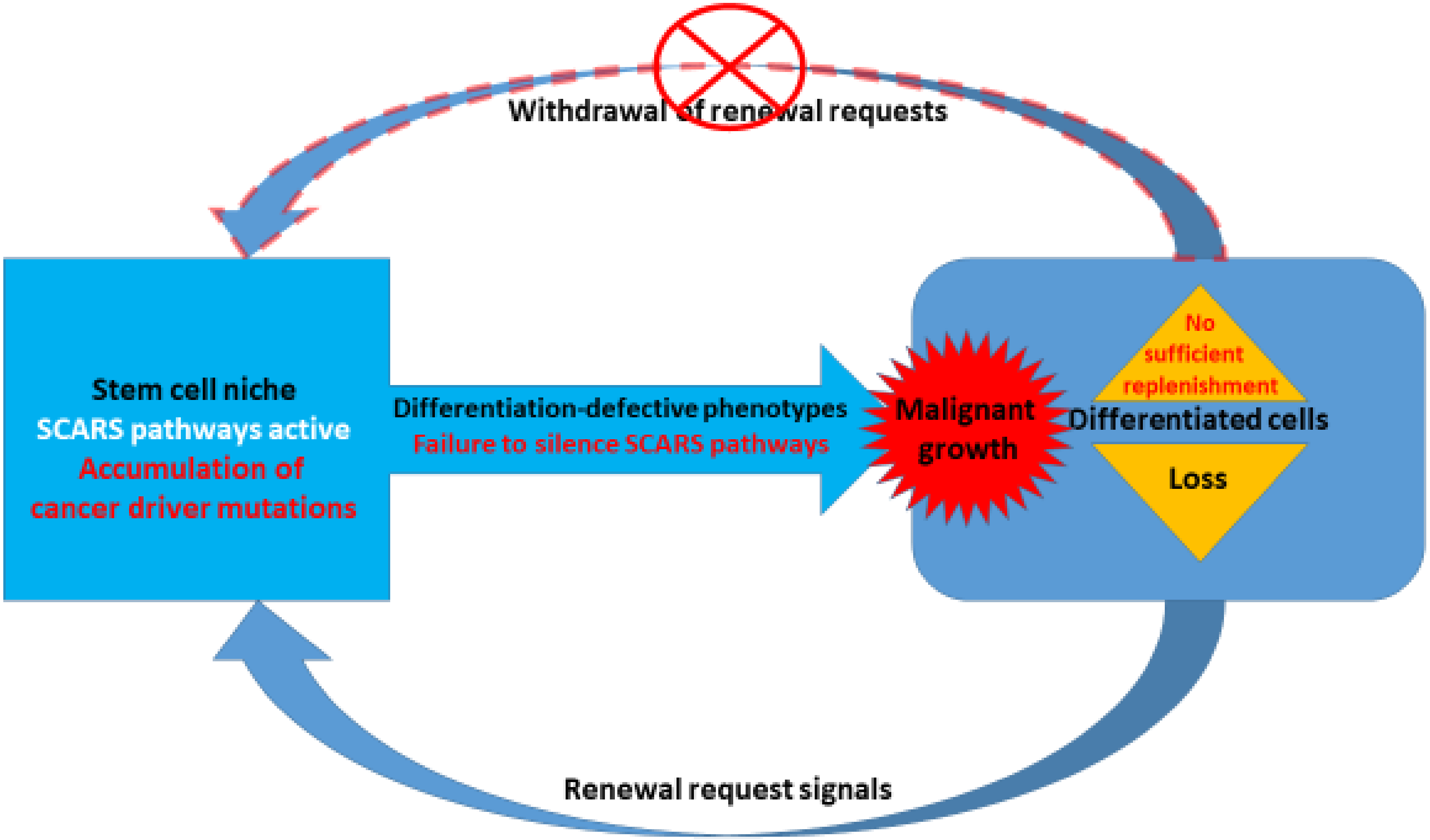

FIGURE 6.